\documentclass[pre,aps,preprint]{revtex4-1}
\usepackage{graphicx}
\usepackage{amsfonts}
\usepackage{amssymb}
\usepackage{amsmath}
\usepackage{textcomp}

\usepackage{color}
\usepackage{psfrag}

\usepackage{epsfig}

\begin{document}

\title{Hyperuniformity, quasi-long-range correlations, and void-space constraints in maximally random jammed
particle packings. II.  Anisotropy in particle shape}

\author{Chase E.~Zachary}
\email{czachary@princeton.edu}
\affiliation{Department of Chemistry, Princeton University, Princeton, New Jersey 08544, USA}
\author{Yang Jiao}
\email{yjiao@princeton.edu}
\affiliation{Department of Mechanical and Aerospace Engineering, Princeton University, Princeton, New Jersey
 08544, USA}
\author{Salvatore Torquato}
\email{torquato@princeton.edu}
\affiliation{Department of Chemistry, Department of Physics, Princeton
Center for Theoretical Science, Program in Applied and Computational Mathematics, and
Princeton Institute for the Science and Technology of Materials,  
Princeton University, Princeton, New Jersey 08544, USA}

\begin{abstract}
We extend the results from the first part of this series of two papers by examining hyperuniformity in 
heterogeneous media composed of impenetrable anisotropic inclusions.  Specifically, we consider
maximally random jammed packings of hard ellipses and superdisks and show that these systems 
both possess vanishing infinite-wavelength local-volume-fraction fluctuations and quasi-long-range 
pair correlations scaling as $r^{-(d+1)}$ in $d$ Euclidean dimensions.  Our results suggest a strong
generalization of a conjecture by Torquato and Stillinger [Phys. Rev. E. \textbf{68}, 041113 (2003)],
namely that all strictly jammed saturated 
packings of hard particles, including those with size- and shape-distributions,
are hyperuniform with signature quasi-long-range correlations.  We show that our arguments concerning 
the constrained distribution of the void space in MRJ packings directly extend to hard ellipse and 
superdisk packings, thereby providing a direct structural explanation for the appearance of 
hyperuniformity and quasi-long-range correlations in these systems.  Additionally, we examine 
general heterogeneous media with anisotropic inclusions and show for the first time that one 
can decorate a periodic point pattern to obtain a hard-particle system that is not hyperuniform with 
respect to local-volume-fraction fluctuations.  This apparent discrepancy can also be rationalized by
appealing to the irregular distribution of the void space arising from the anisotropic shapes of the 
particles.  Our work suggests the intriguing possibility that the MRJ states of hard particles share 
certain universal features independent of the local properties of the packings, including the packing
fraction and average contact number per particle.  
\end{abstract}

\maketitle

\section{Introduction}

In the first part of this series of two papers (henceforth referred to as paper I), we provided a detailed
examination of local-volume-fraction fluctuations in maximally random
jammed (MRJ) packings of polydisperse hard disks.  The MRJ state is defined to be the most disordered 
configuration of impenetrable particles, according to some well-defined order metric, that is 
rigorously incompressible and nonshearable \cite{ToTrDe00}.  Packings of MRJ
monodisperse hard spheres in three dimensions
(3D) have been shown to be hyperuniform, meaning that infinite-wavelength local number density 
fluctuations vanish \cite{ToSt03, DoStTo05, footnoteasymp}.  Additionally, these systems exhibit unusual quasi-long-range 
(QLR) pair correlations
decaying as $r^{-(d+1)}$ in $d$ Euclidean dimensions \cite{DoStTo05}.  Based on the rigidity of the MRJ packings
and the presence of a well-defined contact network, Torquato and Stillinger conjectured that 
any strictly jammed, i.e., 
incompressible and nonshearable, saturated packing of monodisperse hard spheres 
is hyperuniform \cite{ToSt03}, a conjecture for which no counterexample has been found to date.  However,
research into MRJ packings of polydisperse particles has suggested that the conjecture is not true for these 
systems and that QLR correlations are peculiar to the aforementioned monodisperse 
sphere packings \cite{XuCh10, KuWe10}.  

In paper I, we extended
the results of a recent letter \cite{ZaJiTo11} by
presenting definitive evidence that even polydisperse MRJ hard sphere packings in $d$
dimensions are hyperuniform under the more general framework of local-volume-fraction fluctuations, 
whereby one also observes signature quasi-long-range correlations \cite{ZaJiTo10A}.  
In particular, these packings possess asymptotic local-volume-fraction fluctuations decaying faster
than one over the volume of an observation window even though the variance in the 
local number density grows as the volume of the window.
These properties are 
apparently invariant to the degree of polydispersivity and must therefore arise from a fundamental structural
origin in the MRJ packings.  We have argued that maximal disorder of the packings competes with the 
constraints of saturation and strict jamming to homogenize the void space external to the particles while
inducing the observed QLR pair correlations, and our results support a generalization of the Torquato-Stillinger
conjecture that all strictly jammed packings of hard spheres (monodisperse or not) 
in $d$ Euclidean dimensions are hyperuniform with signature QLR correlations.  

\begin{figure}[t]
\centering
$\begin{array}{c@{\hspace{0.6cm}}c}\\
\includegraphics[width=0.45\textwidth]{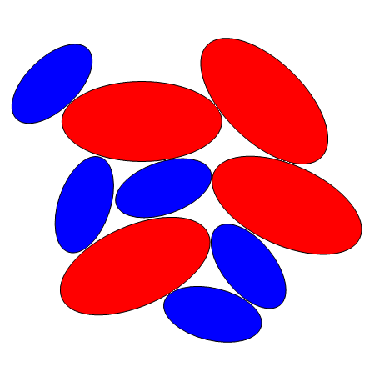}&
\includegraphics[width=0.45\textwidth]{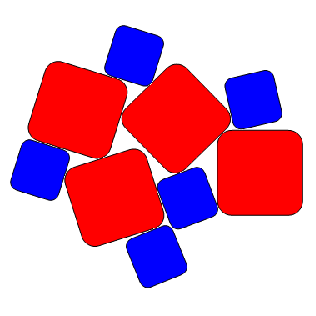}\\
\mbox{\bf (a)} & \mbox{\bf (b)}
\end{array}$
\caption{(Color online)  Illustrative configurations of binary hard ellipses (a) and superdisks (b).  
Excluding particles at the boundary, the packings are locally jammed.  Both
types of particles have an additional rotational degree of freedom not found in sphere packings,
and the anisotropy of the particle shapes has a substantial effect on both the types of local contacts
and the shapes of the local voids between particles.}\label{localfig}
\end{figure}
However, it is not clear that these results should hold for MRJ packings of \emph{nonspherical} 
particles.  Figure 
\ref{localfig} provides local configurations of jammed hard binary ellipses and superdisks (defined below).  
The anisotropy of the particle shapes has a drastic effect on the packing properties of these systems 
compared to hard disks.  Specifically, each particle has an additional rotational degree of freedom, allowing 
for a variety of interparticle contacts.  
The average number of contacts per particle at the MRJ state
is indeed known to be higher for each of these systems compared to hard disks \cite{MaDoStSu05, 
DoCiSaVa04, DoCoStTo07, JiStTo10}, 
and MRJ packings of hard ellipsoids in three dimensions are able to achieve greater packing fractions
than hard spheres \cite{DoCiSaVa04}.  
Particularly for the superdisk packings, the four-fold 
rotational symmetry of the particles favors contacts along the faces \cite{JiStTo08B, JiStTo09B},
but corner-corner and corner-face contacts are also possible
and are common at the MRJ state.  The distortions of the interparticle void shapes are
therefore quite drastic, owing to the complexity of the contact network.  Even 
if our arguments above concerning the 
importance of the void space are true, there is certainly no 
simple reason to believe that these systems
should be hyperuniform at the MRJ state in the generalized sense reported in paper I.  

In this paper we provide direct evidence that, 
despite the aforementioned effects of particle anisotropy
on the properties of the MRJ state, MRJ packings of nonspherical particles are indeed hyperuniform 
with respect to local-volume-fraction fluctuations.  In accordance with
results reported in a recent letter \cite{ZaJiTo11}, we show that these systems also 
have the same types of quasi-long-range correlations decaying as $r^{-(d+1)}$, which
are a likely universal signature of the MRJ state.  Similar QLR correlations have 
also been observed in noninteracting spin-polarized fermion ground states \cite{ToScZa08, ScZaTo09}, 
the ground state
of liquid helium \cite{ReCh67}, and the Harrison-Zeldovich 
power spectrum of the density fluctuations of the 
early Universe \cite{Pe93}, and recent work has provided a direct connection between the presence of 
QLR correlations and the extent of structural order 
in a many-particle system \cite{ZaTo10}.  Our results therefore 
suggest that all of these systems are ``jammed'' in the generalized sense that the local structure is
statistically rigid on the global scale of the system.  We also show that our arguments concerning the
void space of the MRJ packings directly extend to the case of nonspherical particles, thereby providing
a unified explanation for the appearance of hyperuniformity and QLR correlations in these systems.  
It follows that the Torquato-Stillinger conjecture can be further generalized to the 
rather strong statement that
all strictly jammed saturated 
packings of hard particles (spherical or not) are hyperuniform with signature QLR correlations.  

We also examine the converse problem to one considered in paper I and above.  Namely, is it possible 
to construct a general heterogeneous medium from a regular point distribution 
(e.g., a Bravais lattice \cite{FN1}) that is \emph{not} hyperuniform with respect to 
local-volume-fraction fluctuations?  We provide an affirmative answer to this question by
considering regular distributions 
of impenetrable squares in the plane.  By arranging the squares on a square lattice, the anisotropy
of the particles skews the void-space distribution of the medium and prevents infinite-wavelength 
local-volume-fraction fluctuations from vanishing.  However, by examining instead a checkerboard structure,
we are able to recover hyperuniformity by effectively averaging over the induced anisotropy in the void space. 
These results complement our analyses of MRJ packings here and in paper I by emphasizing the 
inherent connection between the void-space distribution and hyperuniformity.  

Section II briefly reviews the concepts of hyperuniformity and jamming as they apply to hard-particle packings.
Section III presents our calculations of local-volume-fraction fluctuations in MRJ packings of 
hard ellipses and superdisks using the methodology discussed in paper I and reviewed in Section II.  
We then consider the effects of particle anisotropy on hyperuniformity in general heterogeneous media in 
Section IV.  Discussion and concluding remarks are in Section V.  A short appendix provides the analysis
necessary for our numerical calculations of the two-point correlations in MRJ hard ellipse and superdisk packings.

\section{Background and Definitions}

\subsection{Hyperuniformity}

Our focus in this work is on local-volume-fraction fluctuations in heterogeneous media.  Formally, a two-phase 
random heterogeneous medium is a region of space partitioned into two distinguishable sets (phases)
$\mathcal{V}_1$ and $\mathcal{V}_2$ with 
interfaces that are known probabilistically  \cite{torquato2002rhm, St95}.  
The fraction of space occupied \emph{globally} by phase $i$
is the \emph{volume fraction} $\phi_i$ of that phase.  However, one can also define a \emph{local volume
fraction} $\tau_i(\mathbf{x})$
as the fraction of space occupied by phase $i$ within some local observation 
region $\mathcal{W}(\mathbf{x}; \mathbf{R})$ with geometric parameters $\mathbf{R}$. 

Unlike the fixed quantity $\phi_i$, 
the local volume fraction fluctuates according to the underlying probability distribution of the 
heterogeneous medium and the location $\mathbf{x}$ of the window.  These fluctuations are 
completely determined by the two-point information of the heterogeneous medium, contained 
within the \emph{two-point probability function} $S^{(i)}_2(\mathbf{r})$, where
\begin{equation}
S^{(i)}_2(\mathbf{r}) = \langle I^{(i)}(\mathbf{r}_1) I^{(i)}(\mathbf{r}_2)\rangle \qquad (\mathbf{r} = \mathbf{r}_2 - 
\mathbf{r}_2).
\end{equation}
The function $I^{(i)}(\mathbf{r})$ is the indicator function for phase $i$:
\begin{equation}
I^{(i)}(\mathbf{r}) = \begin{cases}
1, & \mathbf{r} \in \mathcal{V}_i\\
0, & \mathbf{r} \notin \mathcal{V}_i.
\end{cases}
\end{equation}  
Note that we assume statistical homogeneity of the heterogeneous medium.  

For arbitrary two-phase random media, one can write down successive upper and lower bounds, 
incorporating increasingly higher-order correlation functions,
on the two-point probability function of the void space external to the particles \cite{ToSt83}.  For 
packings of particles, these bounds truncate and become exact at terms involving the pair correlation 
function \cite{FN2}; namely \cite{ToSt85}
\begin{equation}\label{S2imp}
S_2(\mathbf{r}) = 1-2\rho v(\mathbf{R}) + \rho v_{\text{int}}(\mathbf{r}; \mathbf{R}) 
+ \rho^2 (g_2 * v_{\text{int}})(\mathbf{r}; \mathbf{R}),
\end{equation}
where $v(\mathbf{R})$ is the volume of a (possibly anisotropic) particle with geometric parameters $\mathbf{R}$
and $v_{\text{int}}(\mathbf{r}; \mathbf{R})$ is the intersection volume of two such particles with 
centroids separated by a displacement $\mathbf{r}$.  
The corresponding result for the \emph{particle phase} is
\begin{equation}\label{S2p}
S_2^{(p)}(\mathbf{r}) = S_2(\mathbf{r}) - 1+2\rho v(\mathbf{R}),
\end{equation}
containing only two contributions.  
The first term in $S_2^{(p)}(\mathbf{r})$ is related
to the probability of finding two points separated by a displacement $\mathbf{r}$ in the \emph{same}
particle, and the last term accounts for the probability that the points are in separate particles.  It follows 
from these considerations that if a packing has quasi-long-range correlations (as at the MRJ state),
the contribution to $S_2^{(p)}$ from the pair correlation function must be responsible for this behavior.  

The corresponding two-point autocovariance function $\chi(\mathbf{r})$ \cite{torquato2002rhm, St95,
Qu08}
 is obtained by subtracting
the long-range behavior $\phi_i^2$ from $S_2^{(i)}(\mathbf{r})$, and this function is \emph{independent}
of the chosen reference phase $\mathcal{V}_i$, rendering it a global descriptor of correlations within 
the system.  
This property is especially important for MRJ packings because it implies that local-volume-fraction 
fluctuations of the particles and the void space are equivalent, thereby motivating our discussion of the 
fundamental role of the void space in determining hyperuniformity and QLR correlations.  
The Fourier transform $\hat{\chi}(\mathbf{k})$ of 
the autocovariance function is 
known as the \emph{spectral density} \cite{torquato2002rhm}.  
The variance $\sigma^2_{\tau}(R)$ in the local volume fraction is given by \cite{ZaTo09}
\begin{equation}\label{sig}
\sigma^2_{\tau}(R) = \frac{1}{v(R)} \int_{\mathbb{R}^d} \chi(\mathbf{r}) \alpha(r; R) d\mathbf{r},
\end{equation}
where $\alpha(r; R)$ is the intersection volume of two $d$-dimensional spheres of radius $R$ separated
by a distance $r$, normalized by the volume $v(R)$ of a sphere; see Refs. \cite{ToSt03, ToSt06} for exact 
expressions of this function.  For large $R$, the local-volume-fraction variance admits an asymptotic
expansion \cite{ZaTo09}
\begin{equation}
\sigma^2_{\tau}(R) = \frac{\rho}{2^d \varphi}\left\{ A_{\tau} \left(\frac{D}{R}\right)^d + B_{\tau} 
\left(\frac{D}{R}\right)^{d+1} + o\left[\left(\frac{D}{R}\right)^{d+1}\right]\right\},
\end{equation}
where 
\begin{align}
A_{\tau} &= \int_{\mathbb{R}^d} \chi(\mathbf{r}) d\mathbf{r} = \lim_{\lVert\mathbf{k}\rVert\rightarrow 0}
\hat{\chi}(\mathbf{k})\label{Atau}\\
B_{\tau} &= -\frac{\Gamma(1+d/2)}{D\Gamma(1/2)\Gamma[(d+1)/2]} \int_{\mathbb{R}^d}
\lVert\mathbf{r}\rVert \chi(\mathbf{r}) d\mathbf{r}.
\end{align}
The parameter $D$ defines a length scale for the problem (e.g., 
the mean nearest neighbor distance) with a corresponding reduced density,
not necessarily equal to the volume fraction, $\varphi = \rho v(D/2)$. 

It follows from \eqref{Atau} that any heterogeneous medium with a spectral density that vanishes in 
the limit of small wavenumbers possesses asymptotic local-volume-fraction fluctuations decaying 
\emph{faster} than one over the volume of an observation window.  These special systems are known 
as \emph{hyperuniform} \cite{ToSt03, ZaTo09}.  
Since hyperuniform heterogeneous media lack infinite-wavelength
local-volume-fraction fluctuations, the local volume fraction of a reference phase approaches its
global value over relatively few characteristic length scales, implying that the system is globally homogeneous.

In the first part of this series of papers, we expressed the spectral density $\hat{\chi}(\mathbf{k})$
of a finite \emph{hard-particle packing} as a discrete Fourier transform of the local density of particles 
and the particle indicator function $m(\mathbf{r}; \mathbf{R}_i)$, where
\begin{equation}
m(\mathbf{r}; \mathbf{R}_i) = \begin{cases}
1, & \mathbf{r}\text{ is in particle $i$}\\
0, & \text{else};
\end{cases}
\end{equation}
$\mathbf{R}_i$ denotes all geometric parameters of the particle shape.  Specifically, 
we showed \cite{ZaJiTo10A}
\begin{equation}\label{chik}
\hat{\chi}(\mathbf{k}) = \frac{\left\lvert\sum_{j=1}^N \exp(-i \mathbf{k}\cdot\mathbf{r}_j) \hat{m}(\mathbf{k};
\mathbf{R}_j)\right\rvert^2}{V}, \qquad (\mathbf{k} \neq \mathbf{0})
\end{equation}
where $\{\mathbf{r}_j\}$ denotes the particle centroids and $V$ is the volume of the simulation box.
The wavevectors $\mathbf{k}$ are defined by the \emph{dual} lattice vectors to those of the simulation cell;
for example, with a square simulation box of side length $L$, the wavevectors are $\mathbf{k} = (2\pi/L)\mathbf{n}$,
where $\mathbf{n} \in \mathbb{Z}^2$.  Note that the zero wavevector is excluded from the expression \eqref{chik};
we therefore define $\hat{\chi}(\mathbf{0}) \equiv \lim_{\lVert\mathbf{k}\rVert\rightarrow 0} \hat{\chi}(\mathbf{k})$.  

Heterogeneous media with autocovariance functions decaying asymptotically as $r^{-(d+1)}$ exhibit 
anomalous local-volume-fraction fluctuations \cite{ToSt03, DoStTo05, 
ZaTo09, ZaTo10}.  Though still hyperuniform
according to the definition above, these quasi-long-range
pair correlations induce logarithmic corrections in the asymptotic expansion of the local-volume-fraction
variance:
\begin{equation}
\sigma^2_{\tau}(R) \sim \frac{B_0 + B_1 \ln(R)}{R^{d+1}} \qquad (R\rightarrow +\infty).
\end{equation}
In the first paper of this series, we provided strong arguments to indicate that these types of 
quasi-long-range correlations, manifested by a \emph{linear} scaling in the small-wavenumber region
of the spectral density $\hat{\chi}(k)$, are a \emph{signature} of maximally random strictly jammed 
packings of hard $d$-dimensional spheres.  We expand upon those results in this paper by showing that
quasi-long-range pair correlations are likely a universal signature of all MRJ hard-particle packings, including
those packings composed of anisotropic particles.  

\subsection{Jamming in hard ellipse and superdisk packings}

Torquato and Stillinger \cite{ToSt01} have provided a hierarchical classification scheme for jammed hard-particle 
packings, introducing the notions of \emph{local}, \emph{collective}, and \emph{strict} jamming.  Our 
focus in this work is on strictly jammed packings of nonspherical hard particles, in which no global 
boundary-shape deformation accompanied by collective particle motions can exist that respects 
the nonoverlap conditions of the particles.  Strictly jammed packings are therefore rigorously 
incompressible and nonshearable.  

The \emph{maximally random jammed} (MRJ) state is defined to be the most disordered jammed packing,
here assumed to be strictly jammed, according to some well-defined order metrics.  This concept has 
recently replaced the mathematically ill-defined notion of the random close-packed (RCP) state \cite{ToTrDe00}.
Although much work has been done to characterize the structural properties of MRJ hard sphere packings
\cite{ToSt10, ToJi10B, DoStTo05}, 
work has only recently been done to understand MRJ packings of nonspherical particles \cite{JiStTo10}. 
A complete theoretical prediction of the MRJ state is intractable because the problem is inherently nonlocal
with signature quasi-long-range correlations between particles \cite{DoStTo05, ZaJiTo10A}.  
Therefore, methods that 
attempt to study the MRJ state based only on packing fraction and local criteria, such as nearest-neighbor
coordination and Voronoi statistics, are necessarily incomplete \cite{ToSt10}.    
In particular, such local criteria cannot distinguish the MRJ state from other 
``random'' jammed states with higher degrees of order \cite{JiStTo11}.

In this work, we consider binary MRJ packings of hard ellipses and superdisks.  
An ellipse, centered at the origin, is defined by the region
\begin{equation}
\frac{\lvert x_1\rvert^2}{a^2} + \frac{\lvert x_2\rvert^2}{b^2} \leq 1,
\end{equation}
where $2a$ and $2b$ denotes the lengths along each of the semiaxes.  The packing characteristics of 
hard ellipses are determined by the aspect ratio $\alpha = b/a$.  Specifically, it is known \cite{ToSt10,
DoCoStTo07} that 
MRJ packings of hard ellipses and ellipsoids (in three dimensions) are \emph{hypostatic} near the 
sphere point, meaning that the average contact number $Z$ is less than twice the number of 
degrees of freedom per particle.  Even for large aspect ratios, the particles are still slightly 
hypostatic.  Interestingly, MRJ packings of hard ellipsoids in three dimensions have been shown to possess
higher packing fractions than the corresponding MRJ sphere packings \cite{DoCiSaVa04}.  

A superdisk is defined by the region
\begin{equation}
\lvert x_1\rvert^{2p} + \lvert x_2\rvert^{2p} \leq \lambda,
\end{equation}
where $p$ is the so-called deformation parameter that interpolates between the disk $(p = 1)$ and 
square $(p = +\infty)$ shapes, and $2 \lambda^{1/(2p)}$ is the length of the particle along each of its
principle axes.  Unlike ellipses, MRJ superdisk packings are \emph{highly hypostatic}, meaning that
$Z$ is much smaller than twice the number of degrees of freedom per particle for all values of $p$ \cite{JiStTo10}.  
Therefore, to achieve strict jamming, the particles are necessarily correlated in a nontrivial manner; such
correlated structures have been termed \emph{nongeneric} \cite{JiStTo10}.  Interestingly, these nongeneric
structures are not rare configurations \cite{JiStTo10}, reinforcing the notion that local descriptors of the packings
are not sufficient to characterize fully the MRJ state.  

\section{Hyperuniformity in binary MRJ packings of superdisks and ellipses}

In the first part of this series, we clearly demonstrated that ``point'' information contained 
in the distribution of particle centers in polydisperse 
MRJ hard-particle packings is not sufficient to describe local fluctuations appropriately \cite{ZaJiTo10A}.  
Specifically, our results indicate that including the shape information of the particles via
local-volume-fraction fluctuations is essential to 
account for the presence of hyperuniformity and 
quasi-long-range pair correlations in MRJ packings of polydisperse 
hard disks.  However, as previously mentioned, it is not obvious that the arguments from that study 
should apply to MRJ packings of hard anisotropic particles since the types of interparticle contacts
are highly dependent on the particle shape.  

We argued in the case of hard disks
that strict jamming of the packings competes with the maximal randomness of the distribution to 
regularize the void space external to the particles.  Furthermore, the sizes and shapes of the 
voids are inherently correlated with each other over several length scales based on the jamming constraint,
suggesting the presence of signature quasi-long-range pair correlations.  Here we extend the results
of our previous work to MRJ packings of hard ellipses and superdisks and provide
strong evidence for the claim that quasi-long-range correlations
are a \emph{universal} signature of the MRJ state while further supporting our arguments concerning 
the void space distributions of these systems.  

\subsection{Generation of MRJ packings}

\begin{figure}[t]
\centering
$\begin{array}{c@{\hspace{0.6cm}}c}\\
\includegraphics[width=0.45\textwidth]{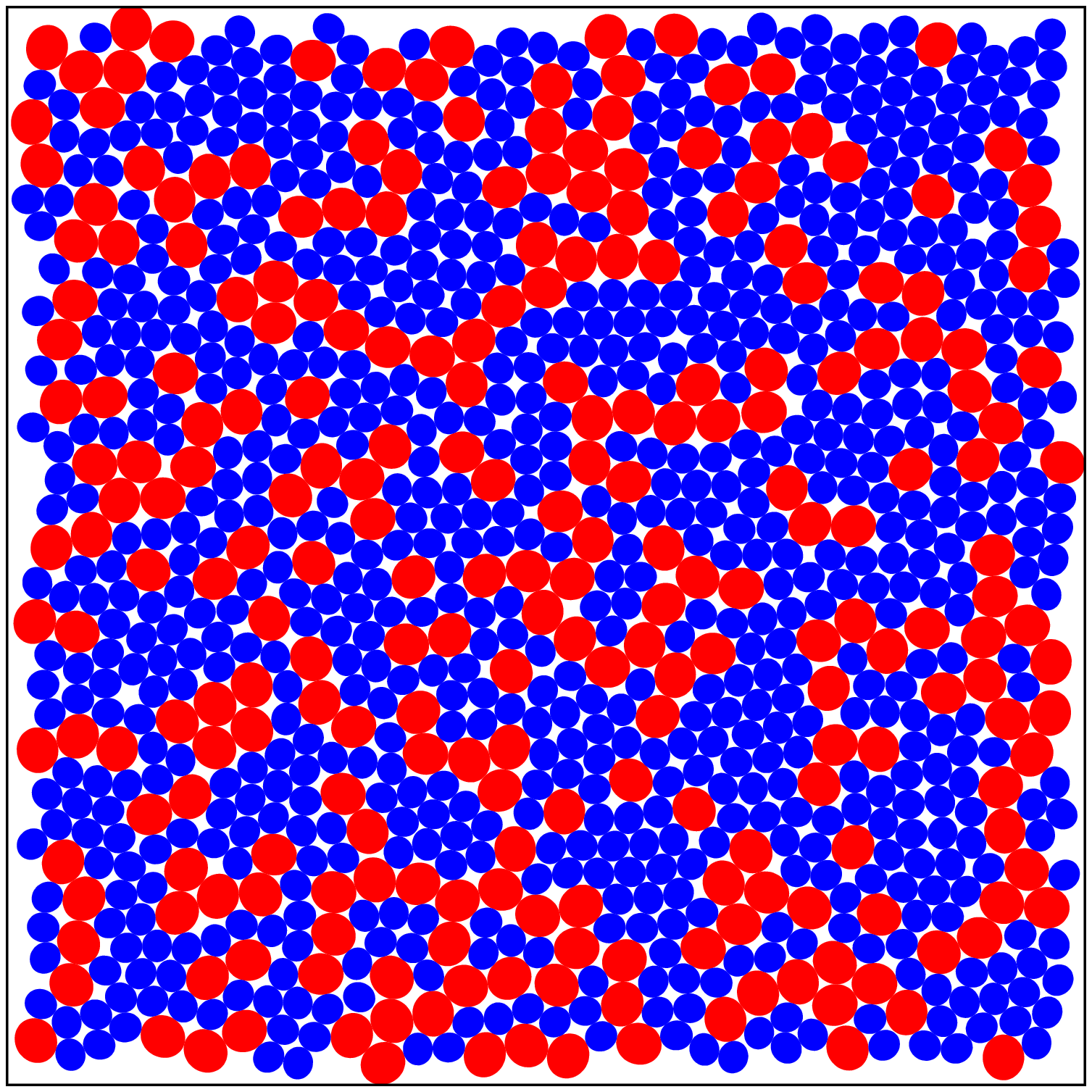} &
\includegraphics[width=0.45\textwidth]{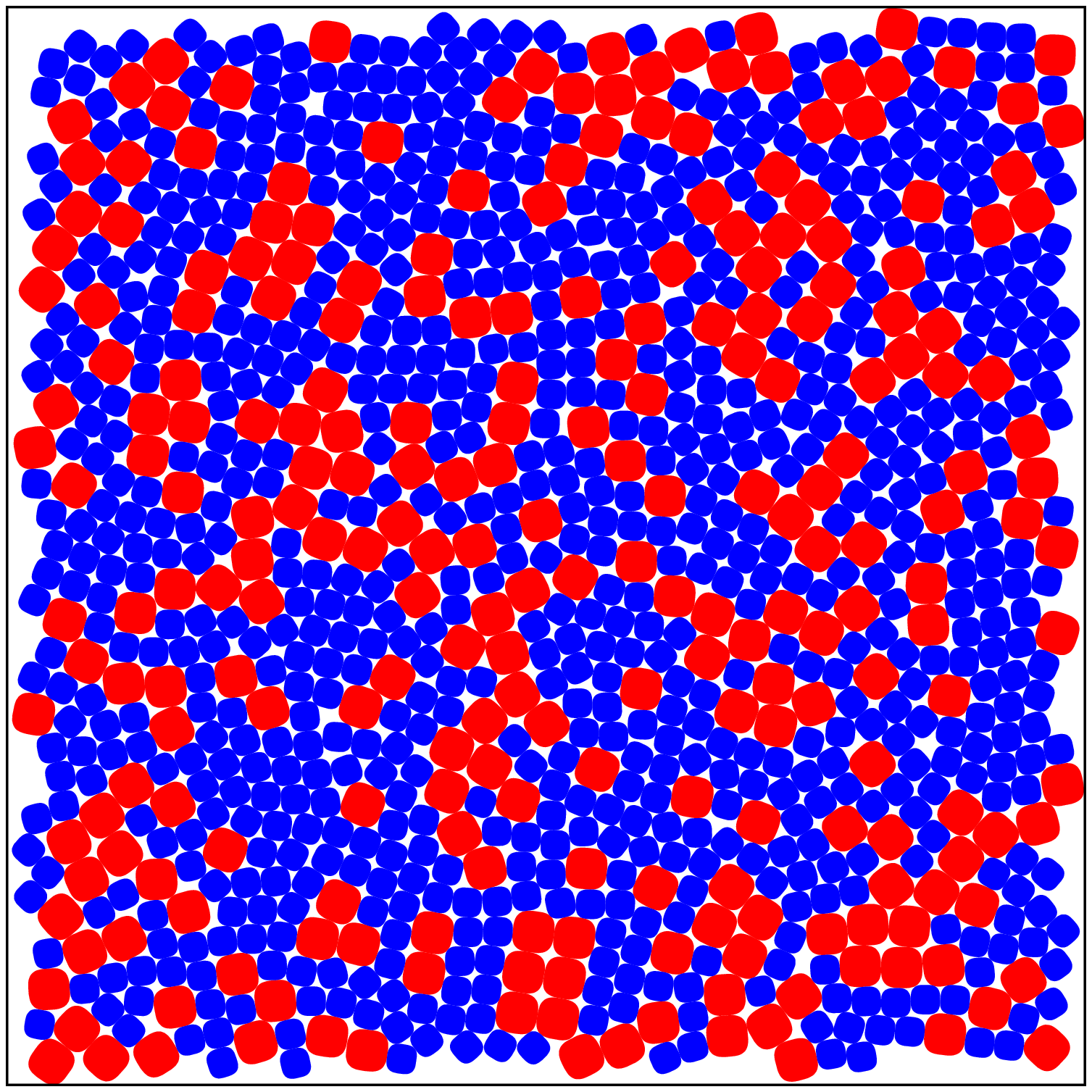} \\
\mbox{\bf (a)} & \mbox{\bf (b)}
\end{array}$
\caption{(Color online)  
(a) Binary MRJ packing of hard ellipses with aspect ratio $\alpha = b/a = 1.1$.  (b)
Binary Packing of hard superdisks with deformation parameter $p = 1.5$ at the MRJ state.  For both configurations,
the size ratio between small (blue) and large (red) particles is $\beta = 1.4$.}\label{configs}
\end{figure}
We have generated configurations of binary MRJ packings of ellipses and superdisks using the 
Donev-Torquato-Stillinger algorithm \cite{DoToSt05}, which is a modified version for
nonspherical particles of the Lubachevsky-Stillinger algorithm \cite{LuSt90, LuStPi91}.  
Particles of two different sizes 
with a fixed size ratio $\beta$ (here $\beta = 1.4$) undergo event-driven molecular dynamics with 
periodic boundary conditions while simultaneously growing at a specified rate.  Toward the end of the simulation,
the unit cell is continuously deformed in order to minimize interparticle gaps, and 
near the jamming point, a sufficiently small expansion rate is used to allow the particles to establish a 
contact network and form an essentially strictly jammed packing.  
The compositions of our packings are $\gamma_{\text{small}} = 
0.75$ and $\gamma_{\text{large}} = 0.25$, where $\gamma_i$ is the mole fraction of species $i$.  Figure
\ref{configs} provides illustrations of our final MRJ packings. 

Although our packings do contain a small fraction of ``rattlers,'' which are particles free to move within some
small cage, the concentration of these rattlers is much smaller than for hard disks.  Indeed, it is 
known that the concentration of rattlers practically vanishes at high aspect ratios for ellipse packings 
\cite{DoCoStTo07} and large deformation parameters for superdisks \cite{JiStTo10}.  
The few rattlers that are present in the packings
must be kept in the final configurations in order to calculate accurately the spectral densities.  
Removing these rattlers introduces large holes into the system, thereby skewing the distribution 
of the void space and breaking hyperuniformity \cite{ZaJiTo10A}.  

\subsection{Structure factor and spectral density calculations}
\begin{figure}[t]
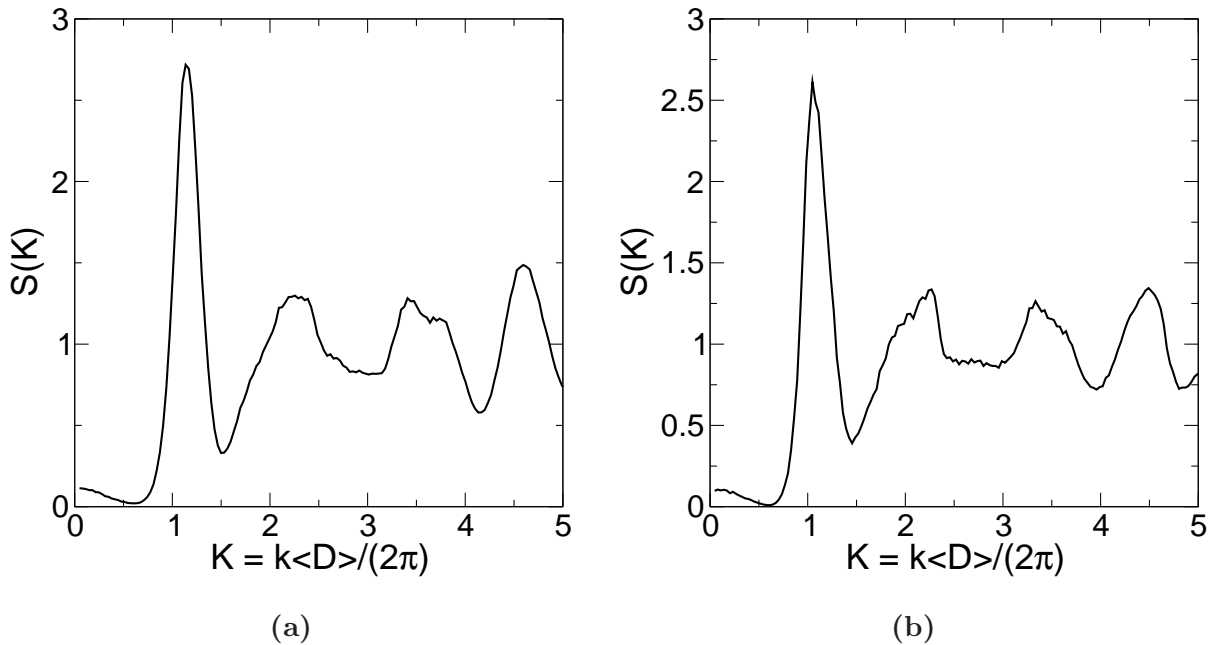

\centering
$\begin{array}{c@{\hspace{0.6cm}}c}\\
\includegraphics[height=3in]{fig3a} &
\includegraphics[height=3in]{fig3b} \\
\mbox{\bf (a)} & \mbox{\bf (b)}
\end{array}$
\caption{Structure factors for binary MRJ packings of hard ellipses with aspect ratio $\alpha = b/a = 1.1$
 (a) and superdisks with deformation parameter $p = 1.5$ (b).  In neither 
case do infinite-wavelength local density fluctuations vanish.}\label{Skfigs}
\end{figure}
Figure \ref{Skfigs} shows the calculated structure factors $S(k)$ for our binary MRJ packings.  The \emph{structure
factor} is related to the pair correlation function $g_2(r)$ between particle centroids according to
\begin{equation}
S(k) = 1+\rho\mathfrak{F}\left\{g_2(r)-1\right\}(k),
\end{equation}
where $\rho$ is the number density of the packing and $\mathfrak{F}$ denotes the Fourier transform.  
The structure factor is related to the fluctuations in the local number density and therefore contains
only ``point'' information of the MRJ packings.  Just as we observed for polydisperse MRJ hard disk packings,
the size distribution of the particles introduces locally inhomogeneous regions of particle centroids such
that infinite-wavelength local number density fluctuations do not vanish, meaning
that $S(0) \neq 0$.  Unlike for packings of hard disks,
the shape anisotropy of ellipses and superdisks compounds this effect since rotations of the particles
increase the types of interparticle contacts that can be formed.  

\begin{figure}[t]
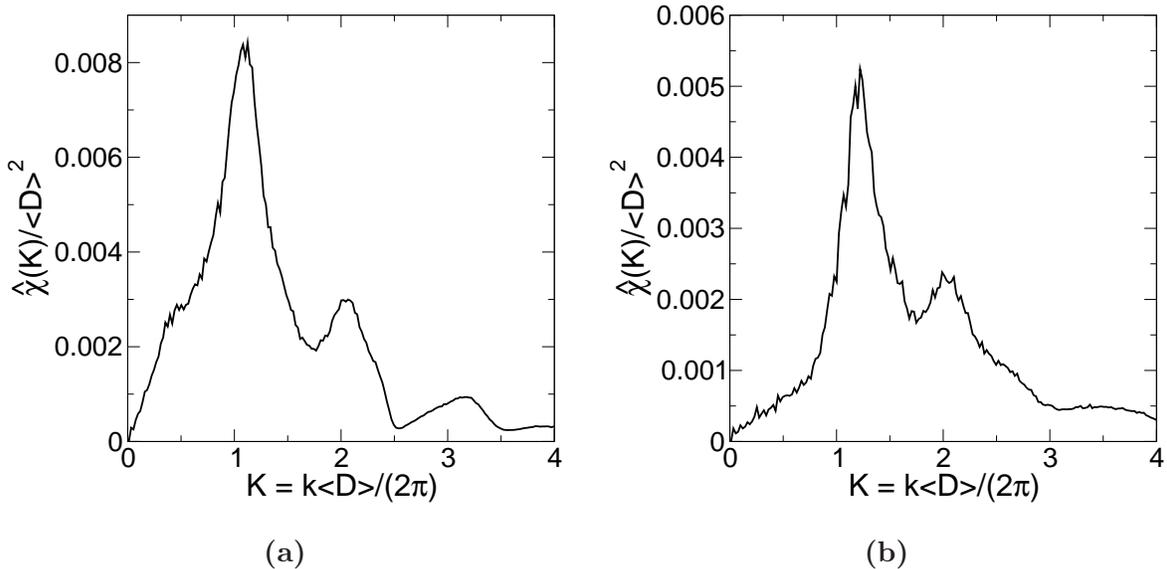

\centering
$\begin{array}{c@{\hspace{0.6cm}}c}\\
\includegraphics[width=0.45\textwidth]{fig4a} &
\includegraphics[width=0.45\textwidth]{fig4b}\\
\mbox{\bf (a)} & \mbox{\bf (b)}
\end{array}$
\caption{Spectral densities for binary hard ellipse packings with aspect ratios $\alpha = b/a = 1.1$ (a)
and $\alpha = 1.4$ (b).  Both systems are hyperuniform with signature 
quasi-long-range correlations.}\label{elchi}
\end{figure}
\begin{figure}[t]
\centering
\includegraphics[width=0.5\textwidth]{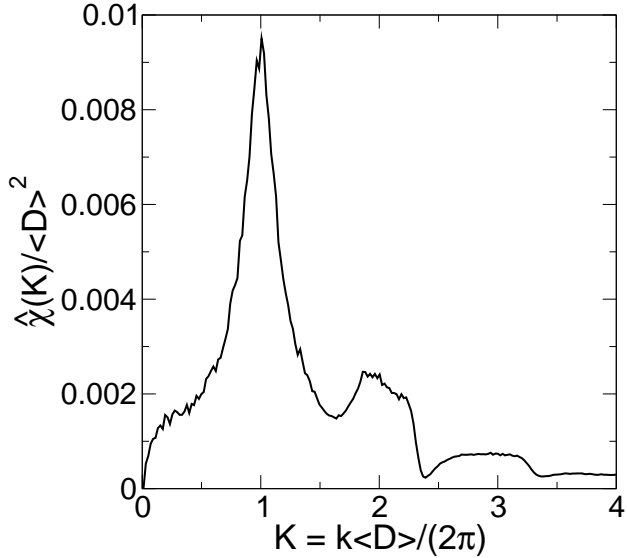}
\caption{Spectral density for a binary hard superdisk packing with deformation parameter $p = 1.5$.}\label{SDchi}
\end{figure}

Using the expression \eqref{chik} for the spectral density and the results in Appendix A for the Fourier transforms
of the particle indicator functions for hard ellipses and superdisks, we have numerically evaluated the 
spectral densities of the corresponding binary MRJ packings, and the results are shown in Figures \ref{elchi}
and \ref{SDchi}.  Both the MRJ ellipse and superdisk packings are indeed hyperuniform with vanishing 
infinite-wavelength local-volume-fraction fluctuations.  Furthermore, in each case we have observed 
a signature linear scaling in the small-wavenumber region of the spectral density, implying the presence of 
quasi-long-range correlations between particles.  This observation is in accordance with our previously-reported
results for MRJ packings of polydisperse disks, but the appearance of the linear scaling in this instance 
is particularly striking since, as we have noted, MRJ packings of anisotropic particles have 
markedly different properties from sphere packings.  Indeed, despite the fact that both the ellipse and 
superdisk packings are hypostatic, the linear scaling persists, suggesting that quasi-long-range correlations
must have a \emph{structural} origin independent of the average contact number.  We have argued 
\cite{ZaJiTo10A} that this essential structural feature is the regularity of the void space external to the particles, 
and we elaborate 
on this point below.

\begin{figure}[t]
\centering
$\begin{array}{c@{\hspace{0.6cm}}c}\\
\includegraphics[width=0.45\textwidth]{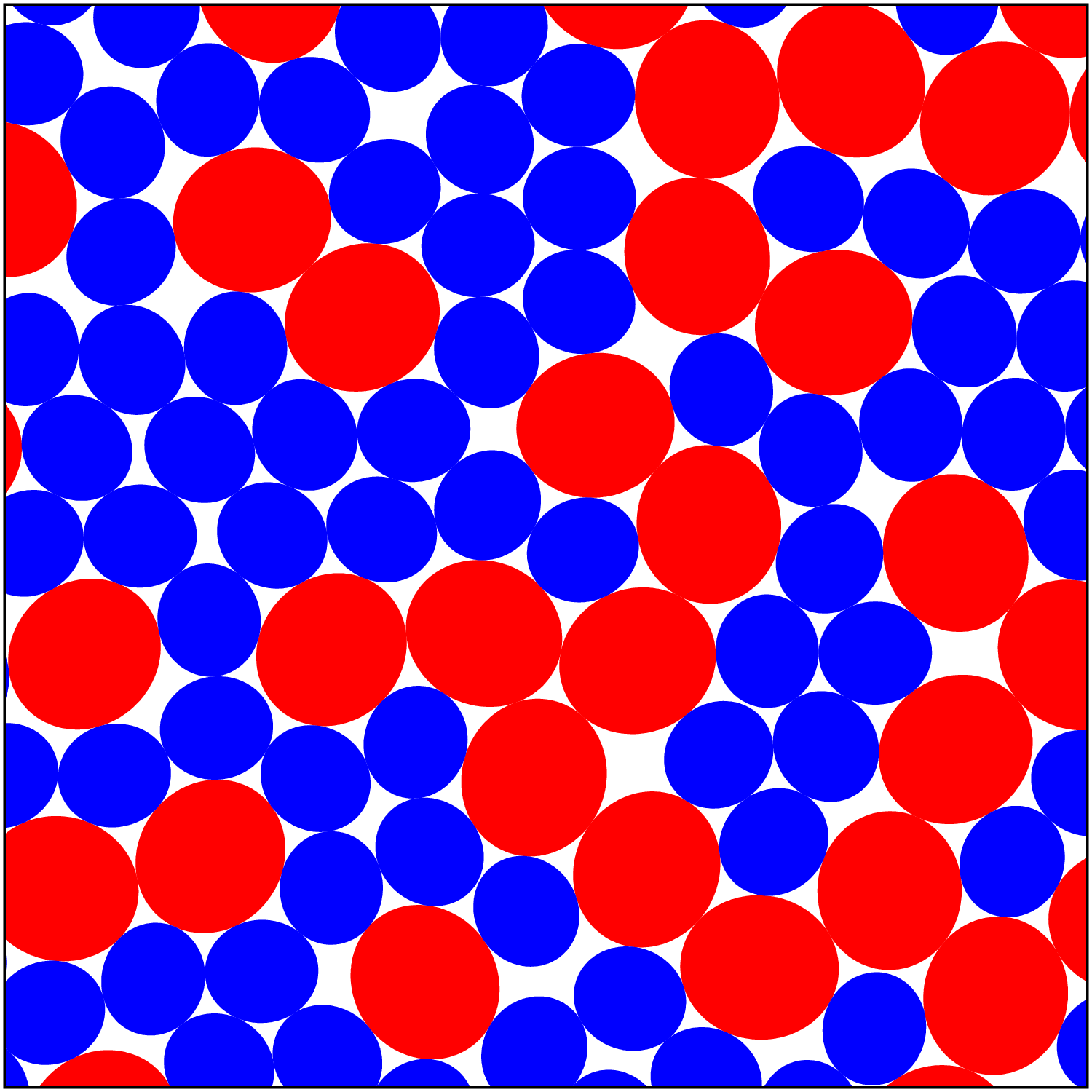} &
\includegraphics[width=0.45\textwidth]{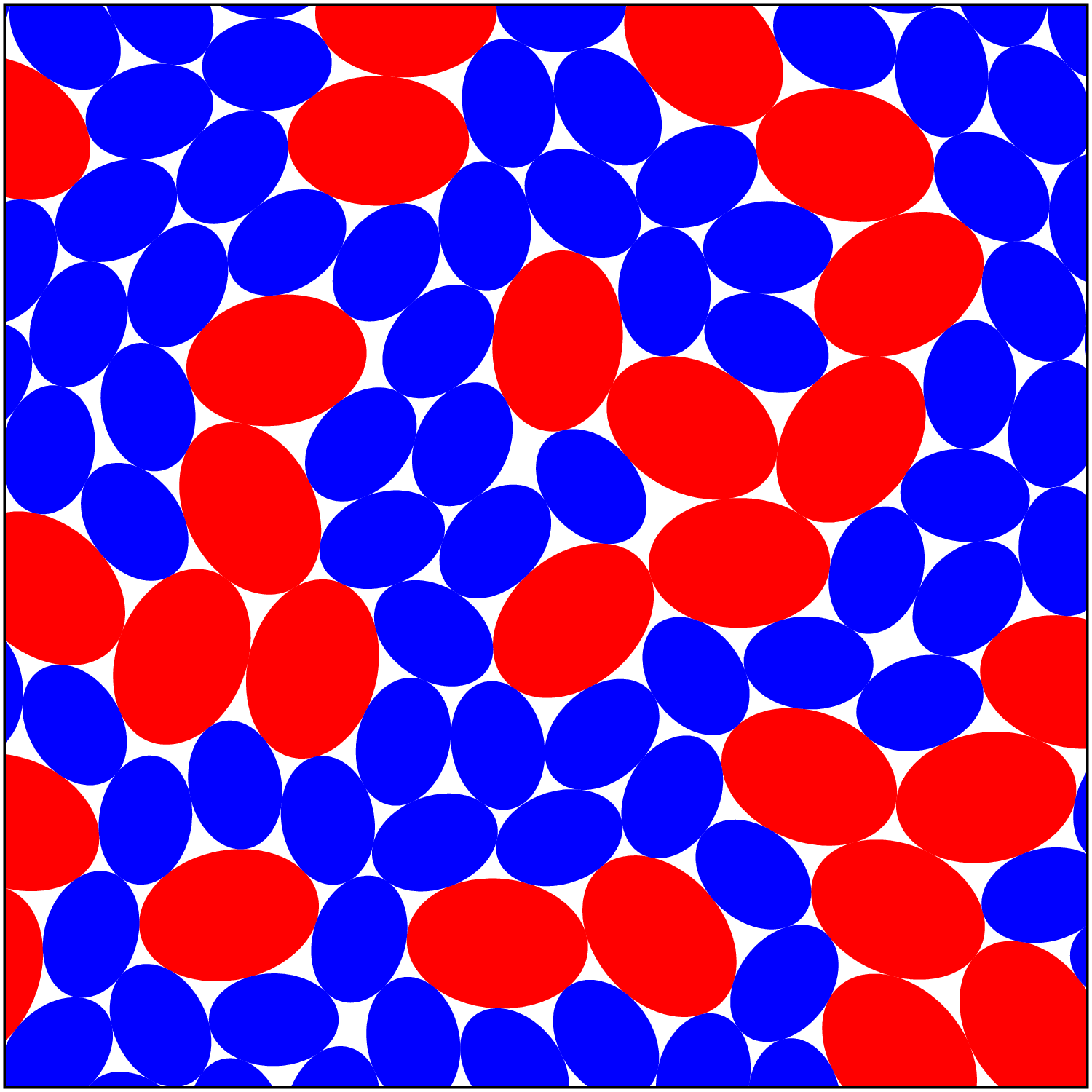} \\
\mbox{\bf (a)} & \mbox{\bf (b)} 
\end{array}$
\caption{(Color online)  Local portions of binary MRJ ellipse packings with aspect ratios $\alpha = 1.1$ (a)
and $\alpha = 1.4$ (b).  By increasing the aspect ratio, the particles increase their average
contact number, thereby homogenizing the void-space distribution.}\label{voids}
\end{figure}

We have also verified in Figure \ref{elchi} that both hyperuniformity and the presence of quasi-long-range 
correlations are invariant to the aspect ratio of our ellipses.  By increasing the aspect ratio, it is known 
that one can increase the density of the MRJ state \cite{ToSt10, DoCiSaVa04, DoCoStTo07} 
along with the average contact number.  The
resulting packing structure is therefore increasingly regularized in the sense that the distribution of 
void shapes and sizes becomes essentially uniform, an effect shown explicitly in Figure \ref{voids}.  
One can immediately see that by increasing the aspect ratio, one increases the number of three-sided
interparticle voids (called $n$-particle loops in paper I)
relative to topologically higher-order voids.  In addition, the average size of the voids becomes
more uniform with increasing aspect ratio, a direct result of the higher average contact number of the particles.  
This increased regularity implies that the void shapes are highly controlled by the jamming constraint,
which is consistent with the presence of quasi-long-range correlations.  

Therefore, although the characteristics of MRJ packings of hard superdisks and ellipses are 
quite distinct from those properties of MRJ disk packings, the presence of hyperuniformity with 
signature quasi-long-range correlations as reflected by a linear small-wavenumber region of the spectral 
density appears to be a universal feature of the MRJ state.  Our arguments concerning the distribution of the 
void space and its relationship to these structural signatures are general enough to account for these 
unexpected properties.  These results suggest that the Torquato-Stillinger conjecture may be substantially 
stronger than previously thought, namely that any MRJ packing of hard particles of \emph{arbitrary} geometry
is hyperuniform with quasi-long-range correlations scaling as $r^{-(d+1)}$ in $d$ Euclidean dimensions.
Although we are still unable to account fully for the origin of the linear small-wavenumber region of the 
spectral density, which, as mentioned in the first part of this series, is tantamount to providing a 
full theoretical prediction of the MRJ state, our work provides a structural explanation for
its appearance by emphasizing the fundamental role of the void space in the packings. 

We remark that one can construct disordered hyperuniform systems with quasi-long-range 
correlations using optimization techniques \cite{UcToSt06}, which do not correspond to 
any strictly jammed packings.  Therefore, it is not our intention to suggest that only 
MRJ packings can be hyperuniform or possess quasi-long-range correlations.  Nevertheless,
such properties are apparently universal among all MRJ packings, regardless of polydispersity 
or particle shape, and our arguments concerning the constrained void space are general enough 
to incorporate even the aforementioned disordered heterogeneous media \cite{ZaTo10}.  


\section{Hyperuniformity and anisotropy in general heterogeneous media}

The binary MRJ packings that we have examined thus far have been statistically homogeneous and isotropic, 
reflecting the symmetry of the chosen decoration for the particle centers.  We have shown 
that it is possible for a point pattern
with non-vanishing infinite-wavelength local number density
fluctuations to generate a \emph{hyperuniform heterogeneous medium} 
by enforcing strict jamming of the system, thereby constraining the available void 
space surrounding the particles in such a way that the pore size becomes effectively uniform.  
Furthermore, this effect is apparently independent of the particle shape.

Our focus is 
now on general statistically \emph{anisotropic} heterogeneous media,
and we provide an example of a system generated by a Bravais lattice
that is not hyperuniform with respect to local-volume-fraction fluctuations.  
This case is therefore the converse problem to the MRJ packings we have considered, namely,
anisotropic decoration of a globally homogeneous point pattern results in a non-hyperuniform
heterogeneous medium.
As before, the reason for
this discrepancy will depend on the distribution of the void space, 
which is made irregular by the anisotropy of the microstructure.  

We recall from \eqref{sig} that anisotropy can affect local fluctuations within a system 
through either the two-point information of the microstructure
or the shape of the chosen observation window via the scaled intersection volume.  In the following analysis we 
consider systems composed of anisotropic inclusions and measure
local volume fractions using an anisotropic observation window.  As has been previously suggested \cite{ToSt03}, 
hyperuniformity of a stochastic system is independent of the shape of the observation window used to measure
local fluctuations.
Specifically, so long as the leading-order
term in the series expansion of the scaled intersection volume $\alpha$ is independent of the size of the window, 
the volume-order term for fluctuations in both the number density and local 
volume fraction is completely determined by the small-wavenumber region of either $S(k)$ or $\hat{\chi}(k)$,
respectively.  
The result is that any two-phase system with a spectral density $\hat{\chi}(k)$ that vanishes 
at the origin
is hyperuniform, and this quantity is independent of the shape of the observation window.  
Therefore, statistical anisotropy of the system plays a fundamental role in 
determining local-volume-fraction fluctuations and is the main focus of this section.

\begin{figure}[!tp]
\centering
$\begin{array}{c@{\hspace{0.6cm}}c}\\
\includegraphics[width=0.45\textwidth]{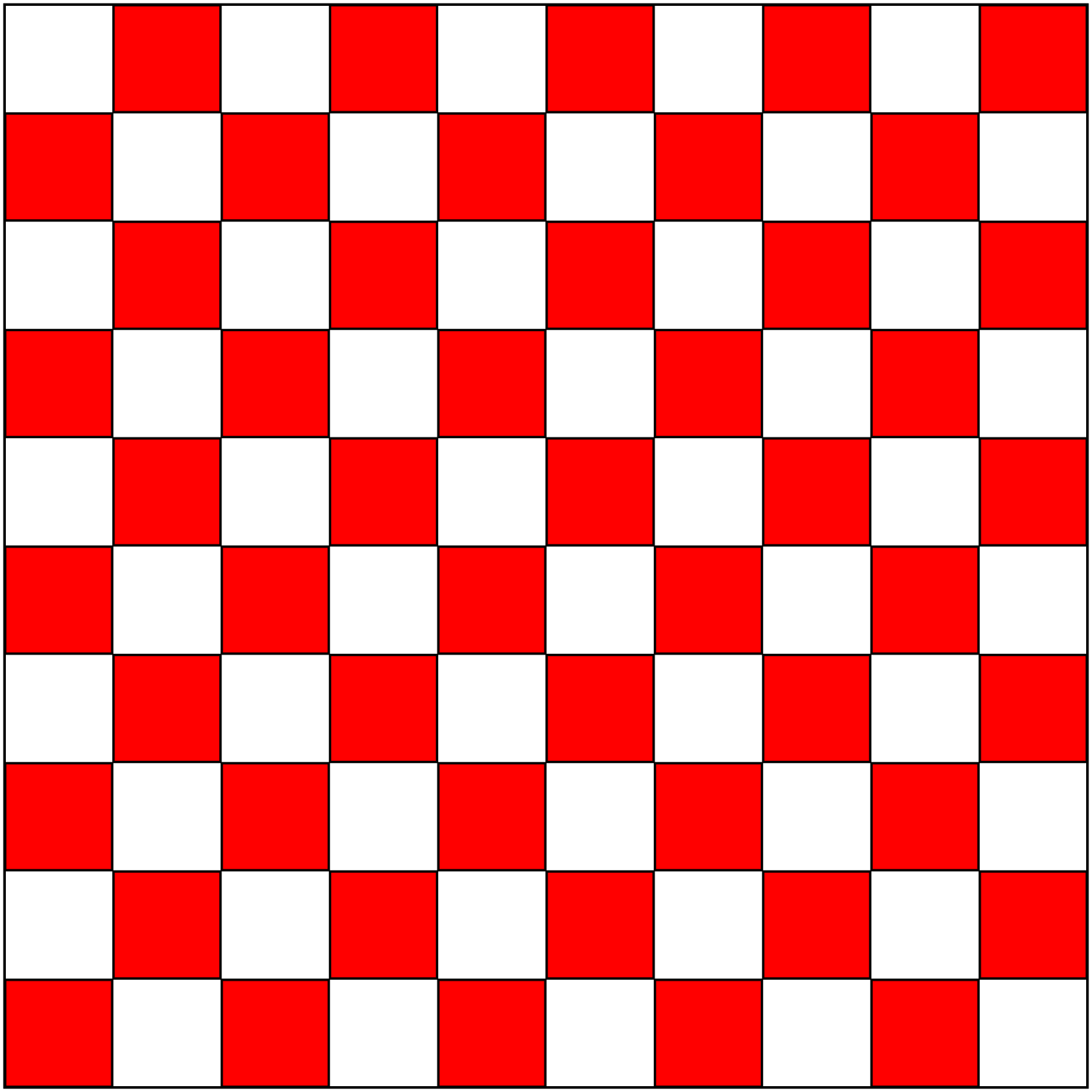} &
\includegraphics[width=0.45\textwidth]{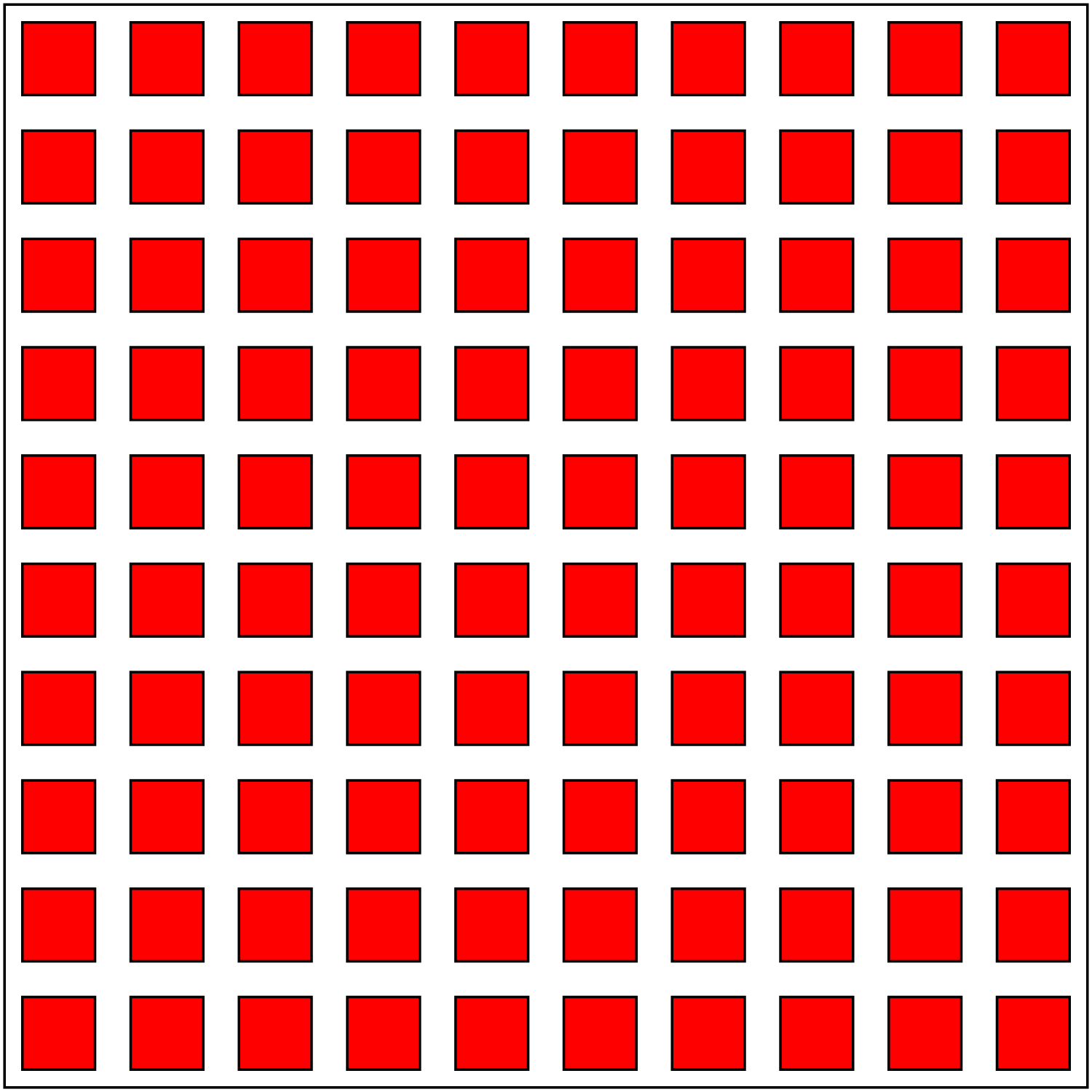} \\
\mbox{\bf (a)} & \mbox{\bf (b)}
\end{array}$
\caption{(Color online)  
(a) The checkerboard pattern. (b)  An anisotropic heterogeneous medium 
obtained by decorating the square lattice with squares; the volume fraction $\phi = 0.5$.}\label{figsix}
\end{figure}

The two systems that we consider here are shown in Figure \ref{figsix}; unlike for the binary 
MRJ packings, these heterogeneous media are composed of nonoverlapping \emph{squares} in the plane.  
The first system shown in Figure \ref{figsix} is a checkerboard 
pattern in which squares of equal size are placed in an alternating manner throughout the plane, 
thereby filling space 
to a volume fraction $\phi = 0.5$.  A related system, 
shown on the right side of Figure \ref{figsix}, is obtained by decorating the square ($\mathbb{Z}^2$) 
lattice with squares; 
this decoration can feasibly be made to obtain any desired volume fraction, and 
without loss of generality we again choose $\phi = 0.5$.  We will henceforth refer to this system as 
the square-$\mathbb{Z}^2$ lattice.

\subsection{Local-volume-fraction fluctuations}

Local-volume-fraction fluctuations in these systems are measured with respect to a square observation 
window of side length $L$.  
We note that the calculation of
the variance $\sigma^2_{\tau}(L)$ requires knowledge of the intersection volume between two 
(oriented) squares of side 
lengths $\ell_1$ and $\ell_2$; this quantity can be
directly evaluated using the definition
\begin{equation}
v_{\text{int}}(\mathbf{r}_1, \mathbf{r}_2; \mathbf{R}_1, \mathbf{R}_2) = \int_{\mathbb{R}^d} 
m(\mathbf{x}, \mathbf{r}_1; \mathbf{R}_1) 
m(\mathbf{x}, \mathbf{r}_2; \mathbf{R}_2) d\mathbf{x},
\end{equation}
where $m(\mathbf{x}, \mathbf{r}; \mathbf{R})$ is the particle indicator function for a point 
$\mathbf{x}$ within an inclusion centered 
at $\mathbf{r}$ with geometric parameters $\mathbf{R}$.  
For a square of side length $\ell$, this indicator function has the form
\begin{equation}
m(\mathbf{x}, \mathbf{r}; \ell) = \Theta\left[\ell-\left(\lvert x_1-r_1\rvert+\lvert x_2-r_2\rvert
+\big\lvert \lvert x_1-r_1\rvert - \lvert x_2-
r_2\rvert\big\rvert\right)\right],
\end{equation}
where $\mathbf{x} = (x_1, x_2)$, $\mathbf{r} = (r_1, r_2)$, and $\Theta$ is the Heaviside step function.  
The calculation of the 
intersection volume for squares can be further 
simplified by noting that the statistics are independent between the orthogonal standard axes for 
$\mathbb{R}^2$.  Given this 
information, one obtains by direct calculation
\begin{align}
v_{\text{int}}(\mathbf{r}_{12}; \ell_1, \ell_2) &= \frac{1}{16}\left(2 x_{12} + \ell_1+\ell_2-\lvert 2x_{12}
+\ell_1-\ell_2\rvert - \lvert 2 x_{12}
-\ell_1+\ell_2\rvert + \lvert -2 x_{12}+\ell_1+\ell_2\rvert\right)\nonumber\\
&\times \left(2 y_{12} + \ell_1+\ell_2-\lvert 2y_{12}+\ell_1-\ell_2\rvert - \lvert 2 y_{12}-\ell_1+\ell_2\rvert 
+ \lvert -2 y_{12}+\ell_1
+\ell_2\rvert\right)\label{fiftyone},
\end{align}
where $\mathbf{r}_{12} = (x_{12}, y_{12}) = (\lvert r_{11} - r_{21}\rvert, \lvert r_{12} - r_{22}\rvert)$.  
It follows from \eqref{fiftyone} and geometric considerations that the intersection volume vanishes when either
$x_{12} \geq (\ell_1 + \ell_2)/2$ or $y_{12} \geq (\ell_1 + \ell_2)/2$.  Note that when $\ell_1 = \ell_2$, 
the result \eqref{fiftyone} simplifies according to
\begin{equation}
v_{\text{int}}(\mathbf{r}_{12}; \ell) = (\ell-x_{12})(\ell-y_{12})\Theta(\ell-x_{12})\Theta(\ell-y_{12})\label{fiftytwo},
\end{equation}
which is simply the product of one-dimensional intersection volumes.  One can now show using \eqref{fiftytwo} 
and the general result \eqref{sig} that 
the variance in the local volume fraction scales with the side length $L$ of a 
square observation window according to
\begin{equation}\label{fiftythree}
\sigma^2_{\tau}(L) = 4\left(\frac{A_{\tau}}{L^2} + \frac{B_{\tau}}{L^3} + \frac{C_{\tau}}{L^4}\right),
\end{equation}
where
\begin{align}
A_{\tau} &=  \int_0^L \int_0^L \chi(x,y) dx dy\label{fiftyfour}\\
B_{\tau} &= - \int_0^L \int_0^L (x+y) \chi(x,y) dx dy\\
C_{\tau} &=  \int_0^L \int_0^L xy \chi(x,y) dx dy.
\end{align}
In the limit of large observation windows, $A_{\tau} \sim \hat{\chi}(0)$ as expected, meaning 
that heterogeneous media for which
local-volume-fraction fluctuations decay faster than one over
the volume of the observation window are hyperuniform \cite{footnotechi}.  
However, unlike for circular observation windows, the scaling in \eqref{fiftythree}
truncates at $\mathcal{O}(L^{-4})$, meaning that 
fluctuations cannot decay faster than one over the square of the observation window volume.  

\begin{figure}[!tp]
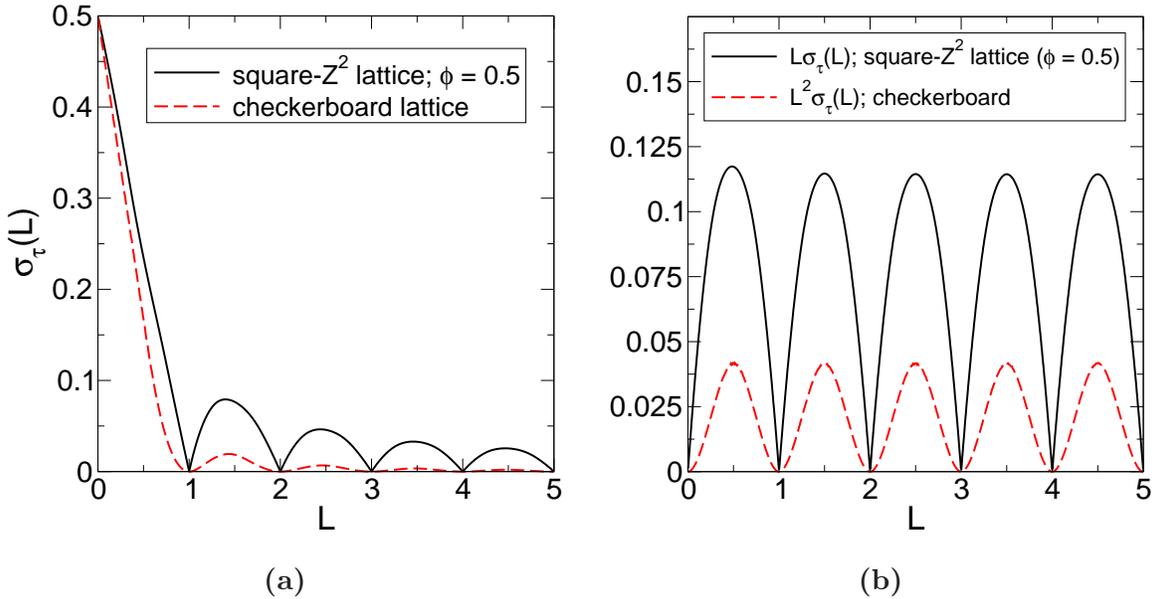

\centering
$\begin{array}{c@{\hspace{0.6cm}}c}\\
\includegraphics[width=0.45\textwidth]{fig8a} &
\includegraphics[width=0.44\textwidth]{fig8b} \\
\mbox{\bf (a)} & \mbox{\bf (b)} 
\end{array}$
\caption{(Color online)
(a)  Fluctuations in the local volume fraction for the checkerboard and square-$\mathbb{Z}^2$ 
heterogeneous media.  
Note that the \emph{standard deviation} and not the variance has been plotted for clarity.
(b)  Leading-order 
terms governing local volume fraction fluctuations for these systems.}\label{figseven}
\end{figure}  
The variance in the local volume fraction for the square-$\mathbb{Z}^2$ 
lattice has been studied \emph{analytically} 
by Quintanilla 
and Torquato \cite{QuTo99} for all volume fractions.  Choosing the side length of an 
inclusion to be $\ell = \sqrt{\phi} < 1$ and defining
\begin{align}
L &= n+\lambda \qquad (\lambda < 1)\\
\delta &= \lvert 1- \ell-\lambda\rvert\\
\Delta &= \lvert \ell-\lambda\rvert\\
m &= [n\ell + \text{max}\{\lambda-1+\ell, 0\}]/L\\
M &= [(n+1)\ell+\text{min}\{\lambda-\ell, 0\}]/L,
\end{align}
one can show \cite{QuTo99}
\begin{equation}
\sigma^2_{\tau}(L) = \frac{[3\delta m^2 + 2L(M^3-m^3)+3\Delta M^2]^2}{9}-[\delta m+L(M^2-m^2)
+\Delta M]^4\label{sixtytwo}.
\end{equation}
The result \eqref{sixtytwo} is plotted in Figure \ref{figseven}. We remark that periodicity of the system 
suppresses 
local-volume-fraction fluctuations when the volume 
of the window is an integral multiple of the lattice spacing.  Quintanilla and Torquato have noted that 
local-volume-fraction 
fluctuations in the square-$\mathbb{Z}^2$ system are indeed 
dampened relative to a random distribution of squares in the plane \cite{QuTo99}; however, the exact 
scaling of the decay 
in the fluctuations has not been previously reported.  
Surprisingly, the right side of Figure \ref{figseven} shows that, while local volume fraction fluctuations 
are suppressed by 
periodicity of the medium, the decay in the 
variance is still controlled by the volume of the observation window; i.e., $\sigma^2_{\tau} \sim 1/L^2$ 
for large windows. 
We have numerically evaluated the asymptotic coefficient $A_{\tau}$ 
(cf. \eqref{fiftyfour}) and obtained $A_{\tau} \approx 0.001805$, which is small but nonvanishing.  
To our knowledge, 
this system is the first example of a 
periodic medium that is not hyperuniform with respect to the variance in the local volume fraction.

We have also numerically investigated the fluctuations in the local volume fraction for the checkerboard 
pattern, and 
the results are shown in Figure \ref{figseven}.  As with the square-integer
microstructure, the checkerboard pattern suppresses local-volume-fraction fluctuations on length scales 
equivalent 
to the size of the unit cell; however, one immediately notes that 
the variance decays to its asymptotic value of zero much more rapidly than in the square-$\mathbb{Z}^2$ system.  
This observation suggests that the checkerboard pattern is indeed hyperuniform,
and Figure \ref{figseven} shows that the decay of the local-volume-fraction variance scales faster than 
the volume of the 
observation window.  In fact, it appears that
the checkerboard pattern actually \emph{saturates} the local-volume-fraction fluctuations, meaning that this system 
exhibits the maximum possible decay  
for a square observation window:  $\sigma^2_{\tau} \sim 1/L^4$.  In other words, it is not only true for this system 
that 
$A_{\tau} = 0$, but also $B_{\tau} = 0$; 
numerical calculations indicate that $C_{\tau} \approx 1.63793 \times 10^{-4}$.

\subsection{Anisotropy and the void space}

In order to understand the differences in the local-volume-fraction fluctuations between these systems, we again 
stress that the shape of the observation window
does not affect volume-order fluctuations in the local volume fraction, meaning that any differences between the 
square-$\mathbb{Z}^2$ and checkerboard systems must be
a result of the spatial distribution of the inclusions themselves.  Specifically, we again focus on the effect of this 
spatial distribution on the available 
void space surrounding the particles.  Both of these systems are statistically anisotropic, meaning that the particle 
inclusions and the void space 
are anisotropic in their spatial distributions.  This claim is readily apparent for the square-$\mathbb{Z}^2$ pattern, 
where the distribution of gaps 
along the standard axes differs from the corresponding distribution along the diagonal.  It is this anisotropy 
that weakens the uniformity of the void space
surrounding the inclusions, thereby preventing the system from achieving hyperuniformity.  

In contrast, the 
symmetry of the void and inclusion phases in the 
checkerboard pattern allows for a more regular spatial distribution even along the diagonals of the system, 
and it is this additional symmetry that 
permits the system to exhibit hyperuniformity with respect to the local volume fraction.  
Indeed, we note that the checkerboard pattern can be generated from 
the square-$\mathbb{Z}^2$ system by rotating each of the particles in the latter microstructure through an angle of 
$\pi/4$ radians.  
It follows that the 
checkerboard pattern overcomes the limitation of anisotropy because it has effectively averaged over the angular 
distribution of the square-$\mathbb{Z}^2$ medium,
thereby enforcing a much stronger constraint on the void space and making it highly regular. Therefore, although 
this example could be considered 
the ``converse'' problem to the MRJ binary packings, it is again the regularity of the void space that controls the 
local-volume-fraction fluctuations.  
Furthermore, we remark that this example highlights the significance of the variance in the local volume fraction 
as a fundamental indicator of 
hyperuniformity in heterogeneous media.  As with the MRJ packings, the information contained in 
the underlying point pattern of the particle centroids
is \emph{not} sufficient to characterize spatial fluctuations resulting from a decoration of the points.  
Even the Bragg peaks occurring in the 
structure factor of a periodic point pattern may not be preserved upon giving the inclusions finite 
volume as with the square-$\mathbb{Z}^2$ microstructure,
thereby breaking hyperuniformity in such instances.  In contrast, local-volume-fraction fluctuations are 
highly sensitive to the homogeneity and isotropy
of the microstructure.

\section{Concluding remarks}

In this series of two papers, we have provided a detailed study of hyperuniformity, jamming, and 
quasi-long-range correlations in MRJ packings of hard particles.  Contrary to previously-published 
work on polydisperse hard-sphere packings \cite{XuCh10, KuWe10}, we have shown that MRJ packings of 
hard particles with both a shape- and size-distribution possess vanishing infinite-wavelength
local-volume-fraction fluctuations and signature quasi-long-range pair correlations.  Our work 
generalizes the Torquato-Stillinger conjecture to the strong statement that all strictly jammed saturated
packings of hard particles are hyperuniform with QLR correlations asymptotically scaling as $r^{-(d+1)}$ 
in $d$ Euclidean dimensions.

We have also identified the first known example of a non-hyperuniform heterogeneous medium obtained 
by decorating an underlying point pattern that possesses vanishing infinite-wavelength local number density
fluctuations.  Our work emphasizes the effect of particle anisotropy on local-volume-fraction fluctuations
and is consistent with our arguments concerning the void space.  By skewing the distribution of voids external
to the particles, one can break hyperuniformity with locally inhomogeneous regions of the microstructure.  
Therefore, a complete description of any heterogeneous material, including granular packings, must account
for the shape information of the particles, and ``point'' information contained in the particle centroids is
insufficient.  

\begin{figure}[t]
\centering
\includegraphics[width=0.5\textwidth]{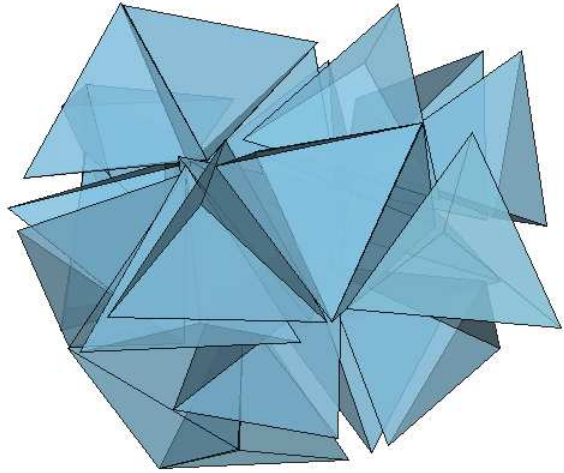}
\caption{(Color online) Local portion of a MRJ packing of tetrahedra.  These systems exhibit the same
signature QLR pair correlations as MRJ packings of spheres, ellipsoids, and superballs
\cite{JiTo10C}.}\label{tetra}
\end{figure}
It is worth mentioning the results of recent work on MRJ packings of the Platonic solids in three dimensions.  
The Platonic solids are convex polyhedra with faces composed of congruent convex regular polygons
\cite{ToJi09}, including the tetrahedron, icosahedron, dodecahedron, octahedron, and cube.  Much attention
has recently been given in the literature to the densest packings of these objects \cite{ToJi09B, ToJi10,
Chen08, ChEnGl10, KaElGr10}, 
but relatively 
little is known about their maximally random jammed structures.  However, current research 
has provided striking evidence that these MRJ packings,
an illustrative portion of which is shown in Figure \ref{tetra}, are indeed hyperuniform with the 
same signature QLR pair correlations \cite{JiTo10C}.  Although in exact accordance with our work, 
these results are striking since these solids each possess sharp corners and edges quite different from 
ellipsoids and superballs, the
three-dimensional generalizations of ellipses and superdisks, respectively.  
Additionally,  of the Platonic solids, only the tetrahedron is 
not centrally symmetric with densest packings that are non-Bravais lattices \cite{ToJi09B, ToJi10}.
Nevertheless, the densest known tetrahedral packings possess a volume fraction much greater 
than spheres.  Very little is currently known about the average contact 
numbers and packing densities of the MRJ states for these solids, and it is therefore not a trivial conclusion
that they should exhibit the same signatures in the MRJ state as the packings in our work.  


{The linear small-wavenumber scaling of the spectral densities of MRJ packings appears to be a universal
feature of these systems, invariant to the particle shape and the size distribution of particles.  We have 
discussed the importance of the homogeneity of the void space external to the particles in promoting 
hyperuniformity and have outlined how correlations between void shapes and sizes may contribute to 
the onset of QLR correlations.  Nevertheless, a complete explanation for the origin of these 
signature correlations is still unavailable, and likely
to remain so in the near future because this problem
is equivalent to quantifying the nature of the MRJ state itself.
Developing such a model is immensely difficult because the problem is inherently non-local, 
meaning that any local analysis is necessarily incomplete \cite{ToSt10}.  Of particular interest in this regard is
recent work suggesting that sub-linear scaling is indeed inconsistent with the jamming and 
impenetrability constraints of the packings \cite{ZaTo10}, 
further supporting our arguments concerning the void space.

\begin{acknowledgments}
This work was supported by the National Science Foundation under Grants DMS-0804431 and 
DMR-0820341.
\end{acknowledgments}

\appendix

\section{Details on the Donev-Torquato-Stillinger algorithm for generating MRJ packings}

As discussed above, we employ the Donev-Torquato-Stillinger molecular dynamics algorithm 
to generate disordered jammed hard-particle packings.  Initially, small nonoverlapping particles 
with a prescribed size distribution and concentration ratio are random placed 
in a simulation box and given random initial velocities.  The system then evolves according to 
Newtonian dynamics as the particles grow with a specific growth rate $\gamma$ under the 
constraint of a fixed size ratio.  To maximize disorder, relatively large growth rates
[i.e., $\gamma \in (0.05, 0.1)$] are initially employed.  Near the jamming point, very small
growth rates are necessary [$\gamma \approx 10^{-6}$] for the particles to establish 
a rigid contact network and strict jamming.  In addition, a deformable simulation box is used 
to facilitate any possible collective motions coupled with boundary deformations, ensuring 
strict jamming of the final configuration.  Jamming is verified by shrinking the particles slightly
and equilibrating the system with a variable simulation box for a sufficiently long time, after which the 
packing is rejammed.  If there is no sufficient structural relaxation, the packing is considered
strictly jammed \cite{DoCoStTo07, JiStTo10}.  
These parameters have been shown to be consistent with the 
translational and orientational order metrics \cite{JiStTo11}.  Statistics for the binary ellipse and 
superdisk packings were obtained by averaging over 20 configurations of 1000 particles each.
We verified that our results were invariant to system size by comparing the calculations with 
systems of up to 10000 particles.

\section{Indicator functions for ellipses and superdisks}

In order to calculate the spectral densities for binary MRJ packings of ellipses and superdisks, it is necessary to 
evaluate the Fourier transforms of the respective particle indicator functions.  Since the analyses for 
these two systems are similar, we discuss the Fourier transform of the ellipse indicator function in 
detail and then state the corresponding result for superdisks.

We recall that an ellipse centered at the origin is defined by the region
\begin{equation}\label{a1}
\frac{\lvert x_1\rvert^2}{a^2} + \frac{\lvert x_2\rvert^2}{b^2} \leq 1,
\end{equation}
where $2a$ and $2b$ denote the lengths of the ellipse along the $x_1$ and $x_2$ semiaxes.  Since 
ellipses in MRJ packings have an additional rotational degree of freedom, we must determine how this 
region is defined upon rotating the ellipse by an angle $\theta$ (counterclockwise) with respect to the 
$x_1$ axis.  Noting that we can always find a coordinate system ($y_1, y_2$) where \eqref{a1}
holds, it follows that the reference frame ($x_1, x_2$) is related to the rotated frame by an orthogonal 
transformation
\begin{equation}
\mathbf{x} = A\mathbf{y},
\end{equation}
where
\begin{equation}
A = \begin{pmatrix}
\cos(\theta) & -\sin(\theta)\\
\sin(\theta) & \cos(\theta)
\end{pmatrix}
\end{equation}
with $\det A = 1$. 
In the reference frame we therefore have the following representation of the ellipse:
\begin{equation}
\frac{\lvert x_1 \cos(\theta) + x_2 \sin(\theta)\rvert^2}{a^2} + \frac{\lvert x_2 \cos(\theta)-x_1\sin(\theta)\rvert^2}{b^2}
\leq 1.
\end{equation}

The particle indicator function $m(\mathbf{x}; a, b, \theta)$ of the ellipse can now be written as
\begin{equation}\label{a5}
m(\mathbf{x}; a, b, \theta) = \Theta\left[1-\frac{\lvert x_1 \cos(\theta)+x_2\sin(\theta)\rvert^2}{a^2}
-\frac{\lvert x_2 \cos(\theta)-x_1\sin(\theta)\rvert^2}{b^2}\right],
\end{equation}
where $\Theta(x)$ is the Heaviside step function.  The Fourier transform of this function is
\begin{equation}
\hat{m}(\mathbf{k}; a, b, \theta) = \int_{\mathbb{R}^2} \exp(-i \mathbf{k}\cdot \mathbf{x}) m(\mathbf{x}; a, b, \theta)
dx_1 dx_2,
\end{equation}
which can be simplified by passing back to the rotated frame and noting that $\det J = 1$, where $J$
is the Jacobian of the transformation.
The result is
\begin{equation}\label{a7}
\hat{m}(\mathbf{k}; a, b, \theta) = \int_{\mathbb{R}^2} \exp(-i \omega\cdot \mathbf{y})
\Theta\left[1-\frac{\lvert y_1\rvert^2}{a^2} - \frac{\lvert y_2\rvert^2}{b^2}\right] dy_1 dy_2,
\end{equation}
where $\omega = A^{\text{T}}\mathbf{k}$. 

The integrals in \eqref{a7} can be evaluated stepwise, first noting that
\begin{align}
I_1 &\equiv \int_{\mathbb{R}} \exp(-i \omega_1 y_1) \Theta\left[1-\frac{\lvert y_1\rvert^2}{a^2}
-\frac{\lvert y_2\rvert^2}{b^2}\right] dy_1\\
&= 2 \int_0^{+\infty} \cos(\omega_1 y_1) \Theta\left[1-\frac{ y_1^2}{a^2}
-\frac{\lvert y_2\rvert^2}{b^2}\right] dy_1 \qquad (\text{symmetry})\\
&= 2\Theta\left[1-\frac{\lvert y_2\rvert^2}{b^2}\right] \int_0^{a \sqrt{1-\lvert y_2\rvert^2/b^2}} \cos(\omega_1 y_1)
dy_1\\
&= 2\Theta\left[1-\frac{\lvert y_2\rvert^2}{b^2}\right] \sin\left(\omega_1 a 
\sqrt{1-\lvert y_2\rvert^2/b^2}\right)/\omega_1.
\end{align}
Substituting this expression into \eqref{a7} gives
\begin{align}
\hat{m}(\mathbf{k}; a, b, \theta) &= (2/\omega_1) \int_{\mathbb{R}} \exp(-i\omega_2 y_2)\Theta\left[1-\frac{\lvert y_2\rvert^2}{b^2}
\right] \sin\left(\omega_1 a \sqrt{1-\lvert y_2\rvert^2/b^2}\right) dy_2\\
&= (4/\omega_1) \int_0^{b} \cos(\omega_2 y_2) \sin\left(\omega_1 a \sqrt{1-y_2^2/b^2}\right)
dy_2 \qquad (\text{symmetry})\label{a14}.
\end{align}
Equation \eqref{a14} can be expressed as a \emph{Laplace convolution} according to
\begin{equation}
\hat{m}(\mathbf{k}; a, b, \theta) = \left(\frac{2 b}{\omega_1}\right) \int_0^1 y^{-1/2} \cos\left(\omega_2 b y^{1/2}\right)
\sin\left(\omega_1 a \sqrt{1-y}\right) dy,
\end{equation}
and this integral can therefore be evaluated analytically using Laplace transforms.  The final result is
\begin{equation}\label{a15}
\hat{m}(\mathbf{k}; a, b, \theta) = 2\pi a b J_1\left(\sqrt{\omega_1^2 a^2 + \omega_2^2 b^2}\right)/\sqrt{\omega_1^2 
a^2 + \omega_2^2 b^2},
\end{equation}
where $J_1$ is the first-order regular Bessel function.  Note that as $\lVert \mathbf{k}\rVert \rightarrow 0$, we
recover the expected result $\hat{m}(\mathbf{0}; a, b, \theta) = v_E(a, b) = \pi a b$,
where $v_E(a, b)$ is the volume of an ellipse with parameters $a$ and $b$.  Also, as $a\rightarrow b$, 
the expression \eqref{a15} reduces to the known result for a disk  \cite{ToSt03}.  

A similar analysis can be done for a superdisk of orientation $\theta$, defined by the region
\begin{equation}
\lvert x_1\cos(\theta)+x_2\sin(\theta)\rvert^{2p} + \lvert x_2\cos(\theta)-x_1\sin(\theta)\rvert^{2p} \leq \lambda,
\end{equation}
where $p$ is the deformation parameter and $\lambda$ determines the diameter of the superdisk along 
one of its principal axes.  The Fourier transform of the particle indicator function for a superdisk can also
be expressed as a Laplace convolution according to
\begin{equation}\label{a17}
\hat{m}(\mathbf{k}; p, \lambda, \theta) = \left(\frac{4\zeta \lambda^{\zeta}}{\omega_1}\right)
\int_0^1 u^{\zeta - 1} \cos\left(\omega_2 \lambda^{\zeta} u^{\zeta}\right) \sin\left[
\omega_1 \lambda^{\zeta} (1-u)^{\zeta}\right] du,
\end{equation}
where we have defined $\zeta = 1/(2p)$.  Although in principle this integral can also be evaluated analytically
with Laplace transforms, for arbitrary $\zeta$ the resulting expressions become intractable.  We therefore 
opt for a series representation of \eqref{a17} that can be efficiently evaluated numerically.   Using the known
expansions
\begin{align}
u^{\zeta-1} \cos\left(\omega_2 \lambda^{\zeta} u^{\zeta}\right) &= 
\sum_{j=0}^{+\infty} \frac{\omega_2^{2j} \lambda^{2\zeta j} u^{\zeta(2j+1)-1}}{\Gamma(2j+1)}\\
 \sin\left[\omega_1 \lambda^{\zeta} (1-u)^{\zeta}\right]/\omega_1 &= \sum_{\ell=0}^{+\infty}
\frac{\omega_1^{2\ell} \lambda^{\zeta(2\ell+1)} (1-u)^{\zeta(2\ell+1)}}{\Gamma[2(\ell+1)]},
\end{align}
we find the result
\begin{equation}
\hat{m}(\mathbf{k}; p, \lambda, \theta) = 4\zeta \sum_{n=0}^{+\infty} \sum_{\ell=0}^{n} \omega_1^{2(n-\ell)}
\omega_2^{2\ell} \lambda^{2\zeta(n+1)} \cdot \frac{\Gamma[\zeta(2\ell+1)]
\Gamma[2\zeta(n-\ell)+\zeta+1]}{\Gamma(2\ell+1)\Gamma[2(n-\ell+1)]\Gamma[2\zeta(n+1)+1]},
\end{equation}
where we have utilized the \emph{Cauchy product} of two infinite series in order to facilitate 
a numerical evaluation of the final result.  One can immediately verify that 
\begin{equation}
\hat{m}(\mathbf{0}; p, \lambda, \theta) = v_S(p, \lambda) = (2/p)\lambda^{1/p} B[1/(2p), 1+1/(2p)],
\end{equation}
where $v_S(p, \lambda)$ is the volume of a superdisk and $B(x, y) = \Gamma(x)\Gamma(y)/\Gamma(x+y)$
is the beta function.  Note also that as $p\rightarrow 1$, $\hat{m}(\mathbf{k}; p, \lambda, \theta)$ 
converges to the known result for a circular disk.


\begin{thebibliography}{38}
\bibitem{footnoteasymp}
Vanishing infinite-wavelength local-number-density or local-volume-fraction fluctuations
  implies that zero-wavenumber Fourier modes possess zero amplitude.
  This concept should be contrasted with ``asymptotic'' fluctuations, which grow 
  more slowly than the volume of an observation window for the number-variance problem
  and decay to zero faster than the reciprocal volume for local-volume-fraction fluctuations.
\bibitem{FN1}
  A Bravais lattice in $d$ Euclidean dimensions is a periodic point pattern generated by
  taking integer linear combinations of a set of $d$ basis vectors $\{\mathbf{e}_i\}$.
\bibitem{FN2}
  The pair correlation function $g_2(r)$ is defined such that $\rho s(r) g_2(r) dr$
  is the average number of particles within a spherical shell, centered
  on a particle at the origin, of radius $r$ and thickness $dr$.
\bibitem{footnotechi}
Hyperuniformity in heterogeneous media 
is defined by the 
asymptotic behavior of the local-volume-fraction fluctuations.  Although
this definition can be translated into a condition on the spectral density,
this condition does not in turn necessarily define hyperuniformity.  
For the example of the square-$\mathbb{Z}^2$ lattice considered in the text,
anisotropy of the particles implies that one should examine the full 
anisotropic spectral density, which we avoid by directly calculating 
the local-volume-fraction fluctuations.
\end{thebibliography}

\end{document}